\documentclass[aps,showpacs,manuscript,12pt]{revtex4}
\usepackage{amssymb}
\usepackage{amsmath}
\usepackage{graphicx}

\begin{document}

\title{\textbf{Modelling of anthropogenic pollutant diffusion in the atmosphere and
applications to civil protection monitoring$^{\S }$}}
\author{Marco Tessarotto$^{a,b,c}$
and Massimo Tessarotto$^{c,d}$} \affiliation{\ $^{a}$Civil
Protection Agency, Regione Friuli Venezia-Giulia, Palmanova,
Italy, $^{b}$Department of Electronics, Electrotechnics and
Informatics, University of Trieste, Italy, $^{c}$Consortium of
Magneto-fluid-dynamics, University of Trieste, Italy,
$^{d}$Department of Mathematics and Informatics, University of
Trieste, Italy}

\begin{abstract}
A basic feature of fluid mechanics concerns the frictionless
phase-space dynamics of particles in an incompressible fluid. The
issue, besides its theoretical interest in turbulence theory, is
important in many applications, such as the pollutant dynamics in
the atmosphere, a problem relevant for civil protection monitoring
of air quality. Actually, both the numerical simulation of the ABL
(atmospheric boundary layer) portion of the atmosphere and that of
pollutant dynamics may generally require the correct definition of
the Lagrangian dynamics which characterizes arbitrary fluid
elements of incompressible thermofluids. We claim that
particularly important for applications would be to consider these
trajectories as phase-space trajectories. This involves, however,
the unfolding of a fundamental theoretical problem up to now
substantially unsolved: {\it namely the determination of the exact
frictionless dynamics of tracer particles in an incompressible
fluid, treated either as a deterministic or a turbulent (i.e.,
stochastic) continuum.} In this paper we intend to formulate the
necessary theoretical framework to construct such a type of
description. This is based on a phase-space inverse kinetic theory
(IKT) approach recently developed for incompressible fluids
(Ellero \textit{et al.}, 2004-2008). {\it Our claim is that the
conditional frictionless dynamics of a tracer particles - which
corresponds to a prescribed velocity probability density and an
arbitrary choice of the relevant fluid fields - can be exactly
specified}.
\end{abstract}

\date{\today }
\maketitle


\section{Introduction}

This work (together with Refs.\cite{Tessarotto2008}) is a part of
a research project related to the theoretical description of
pollutant dynamics in the atmosphere, a subject relevant for civil
protection monitoring of air quality in the atmosphere. Here, we
refer in particular to the so-called ABL (atmospheric boundary
layer) portion of the atmosphere where the earth surface (land or
water) has a direct influence and most of pollution releases
occur. In fact, the ability\ to predict the dynamics of
anthropogenic pollutants, especially in order to estimate their
concentration at ground level, is a prerequisite for environmental
investigations. A critical issue is therefore the identification
of mathematical models, able to give reliable predictions for
pollutants concentrations in the presence of complex terrain and
for prescribed weather profiles. This involves the ability to
simulate pollutant dynamics in a variety of different physical
conditions. In fact in the ABL, sufficiently close to the ground,
the atmosphere is mainly characterized by a turbulent flow arising
from the wind shear produced by friction with the ground surface.
Instead, at the top of the ABL, in the free atmosphere, the wind
speed is approximately geostrophic and therefore possibly laminar.
The stronger the wind, the more intense is the generated
turbulence arising close to the ground, whose properties may be
very different, depending on the vertical temperature gradient of
the atmosphere. Since turbulence reinforces mixing it tends to
homogenize the fluid much more quickly than would a laminar flow,
thus preventing local accumulation of pollutants. In addition
meteorological parameters (of the atmosphere) are strongly
affected by the earth's surface through dynamical processes
(friction of the air over the surface and through thermal
processes heating or cooling of the air in contact with the
ground). Until recently, fully reliable mathematical methods of
this type, able to take into account the full complex
phenomenology of the atmosphere and simulate the dynamics of
particle or gaseous pollutants in the atmosphere, have been
missing. For this reason in the past, most of predictions for
pollutant transport in the atmosphere have been based on
wind-tunnel experiments. Purpose of the presentation is an
overview of the mathematical models currently available and a
brief analysis of new theoretical developments in the field. In
particular, here we intend to refer to the statistical approach
developed by Marco Tessarotto et al. \cite{Tessarotto2008} based
on the IKT (see also Ellero et al., 2000-2008,
\cite{Ellero2000,Ellero2004,Ellero2005,Tessarotto2006,Tessarotto2007,Tessarotto2008-aa}).
In the sequel, we shall concentrate on those issues that we
consider mostly relevant for the present investigation. We intend
to show that the IKT approach is a useful theoretical framework
that can be used to simulate the dynamics of pollutants in
prescribed fluid flows, taking into account, besides the velocity
and pressure fields also the temperature of the same fluid. The
mathematical approach appears therefore susceptible of actual
applications for simulations of pollutants dynamics in the
atmosphere, such as those relevant to civil protection monitoring.

\section{The present state of research}

The subject of the investigation (pollutant dynamics) involves
some of the most disputed and still open issues in fluid dynamics.
The motion of particles in a fluid as they are pushed along
erratic trajectories by fluctuations of the fluid fields is
fundamental to transport and mixing in
turbulence. It is essential, for example, in combustion processes \cite%
{Pope1994}, in the industrial production of nanoparticles \cite%
{Pretsinis1996} as well as in atmospheric transport, cloud
formation and air-quality monitoring of the atmosphere
\cite{Villancourt2000,Weil1992}. The Lagrangian approach - denoted
as Lagrangian turbulence (LT) - has been fruitful in advancing of
the understanding of the anomalous statistical properties of
turbulent flows \cite{Falkovich2001}. In particular, the dynamics
of particle trajectories has been used successfully to describe
mixing and transport in turbulence \cite{Pope1994,Shraiman2000}.
Nevertheless, issues of fundamental importance remain unresolved.
\ As it is well known, the description of fluids can be performed
choosing either Eulerian or Lagrangian viewpoints. The two
approaches are equivalent and, if fluid dynamics were fully
understood, one should be able to translate Eulerian properties
into Lagrangian ones and viceversa. We are still quite far from
this goal.

In the past the treatment of Lagrangian dynamics in turbulence was
based on stochastic models of various nature, pioneered by the
meteorologist L. F. Richardson \cite{Richardson}. These models,
which are based on tools borrowed from the study of random
dynamical systems, typically rely - however - on experimental
verification rather than on first principles. A first example is
provided by refined stochastic models - based in non-extensive
statistical mechanics \cite{Beck2001,Mordant2004} - which try
to reproduce the observed intermittency of turbulence \cite%
{Sawford2003,Reynolds2003}. Another approach is based on
Brownian-motion-type models, which involve stochastic particle
equations of motion and an associated Fokker-Planck equation (see,
for example, Subramanian et al.,2004 \cite{Subramanian2004}). In
such models the diffusion tensor in the Fokker-Planck equation may
depend on the local velocity shear rate, implying a possible
violation of the classical (equilibrium) fluctuation-dissipation
theorem \cite{Santamaria2001}. A further class of models has been
based on kinetic theory, enabling us to obtain the Fokker-Planck
equation for the motion of Brownian particles in rarefied
nonuniform gases. In particular, by expanding in term of the
mass-ratio (of the light-gas molecular mass to the
Brownian-particle mass) the Boltzmann collision operator, the
light-gas distribution function was approximated by the first two
terms in the Chapman-Enskog expansion. As a result, it was found
to obey a Fokker-Planck equation with a diffusion tensor
independent of the light-gas velocity gradients
\cite{DelaMora1982, Rodriguez1983}.

However, in most cases there remains a lack of experimental data
to verify the reliability of such models. Verification can be
based, in particular, on the measurement of fluid particle
trajectories, obtained by seeding a turbulent flow with a small
number of tracer particles and following their motions with an
imaging system. \ On the other hand, the accurate evaluation of
the Lagrangian velocity in laboratory turbulence experiments
requires measurements of positions of tracer particle by using a
suitable tracking system able to resolve very short time (and
spatial) scales. In practice this can be a very challenging task
since particle motions must be measured
on very short time scales of the order of the Kolmogorov time,%
\begin{equation}
\tau =(\nu /\varepsilon )^{1/2}
\end{equation}%
where $\nu $ is the kinematic viscosity and $\varepsilon $ the
turbulent energy dissipation. Similarly, to get the Lagrangian
acceleration one should have experimental access to time scales
again comparable or even smaller than the Kolmogorov time scale of
the flow. Another important physical observable is the pressure
(and the related pressure gradient), which is typically hard to
measure experimentally.\ Despite these difficulties, growing
interest in studying Lagrangian turbulence is motivated by the
recent advances in laboratory and numerical experiments \cite%
{LaPorta2001,Gotoh2002,Biferale2004}). In particular, the role of
numerical simulations represents a new challenge, since they are
expected to represent alternative tools to laboratory experiments.

As for the theory itself, rigorous results have been scanty,
probably because of the subject complexity. \textit{In particular,
an open issue is the very definition of the dynamics of tracer
particles which may injected in the fluid with arbitrary initial
velocities. These velocities in practice may be generally very
different from the local fluid velocity.} In contrast, in
customary treatments the dynamics of tracer particles is
identified with that of the fluid particles, which is determined
uniquely by the Navier-Stokes equations. As a consequence,\emph{\
}the dynamics (i.e., in particular the acceleration) of tracer
particles in incompressible thermofluids might be very different
from that of the fluid elements. This problem is not only a
challenging theoretical issue, but is obviously of primary
importance for its implications in computational and environmental
fluid dynamics.

In this connection, however, recently there has been an important
breakthrough, represented by the discovery of the so-called
inverse kinetic
theory (IKT) approach for incompressible fluids \cite%
{Ellero2000,Ellero2004,Ellero2005,Tessarotto2006,Tessarotto2007}.
In accompanying papers \cite{Cremaschini2008,Tessarotto2008} the
same approach has been extended to the treatment of incompressible
thermofluids.

\textit{In this paper we claim that the IKT of
Ref.\cite{Tessarotto2008} permits us to determine uniquely the
(frictionless) dynamics of tracer particles}.

In fact, a key feature of the this type of approach is that\ it
affords a straightforward connection between Eulerian and
Lagrangian fluid descriptions. This is achieved by identifying the
relevant fluid fields,
which are assumed to be defined in a suitable domain $\Omega \subseteq $ $%
\mathbb{R}
^{3}$ (fluid domain), with appropriate moments of a
suitably-defined kinetic distribution function density
$f(\mathbf{x,}t)$ [with $\mathbf{x=(r,v)}\in \Gamma \mathbf{,}$
$\mathbf{x}$ and $\Gamma $ denoting a suitable state-vector and an
appropriate phase-space] which is assumed to advance in time by
means of Vlasov-type kinetic equation. In such a case, the
time-evolution of the kinetic distribution function is determined
by a kinetic equation which, written in the Eulerian form, reads
\begin{equation}
Lf(\mathbf{x},t)=0.
\end{equation}%
Here $L$ is the streaming operator $Lf\equiv \frac{\partial }{\partial t}f+%
\frac{\partial }{\partial \mathbf{x}}\cdot \left\{ \mathbf{X}(\mathbf{x}%
,t)f\right\} $ and $\mathbf{X}(\mathbf{x},t)\equiv \left\{ \mathbf{v,F}(%
\mathbf{x},t)\right\} $ a suitably smooth vector field, while
$\mathbf{v}$ and\textbf{\ }$\mathbf{F}(\mathbf{x},t)$ denote
respectively appropriate velocity and acceleration fields. As a
main consequence the approach can in principle be used to
determine in a rigorous way the Lagrangian formulation for
arbitrary complex fluids. Although the choice of the phase space
$\Gamma $ is in principle arbitrary, in the case of incompressible
isothermal fluids, it is found \cite{Ellero2004} that the
phase-space $\Gamma $ can
always be reduced to the direct-product space $\Gamma =\Omega \times V$ (%
\emph{restricted phase-space})$,$ where $\Omega ,V\subseteq $ $%
\mathbb{R}
^{3},$ $\Omega $ is an open set denoted as configuration space of
the fluid (fluid domain) and $V$ is the velocity space. \ This
type of approach (based on a restricted phase-space IKT
formulation) will be adopted also in the sequel. The main
motivation [of this work] is that some of the general
understanding recently achieved in simple flows by means of the
IKT approach could also give a significant contribution to a wider
range of problems. In the sequel, we will concentrate on the issue
of a consistent Lagrangian formulation for fluid dynamics based on
a \emph{phase-space (IKT) description of incompressible fluids,
}whereby its pressure, velocity (and possibly also thermal)
fluctuations are consistently taken into account in
the phase-space dynamics of suitable \textit{phase-space inertial particles}%
, i.e., particles whose dynamics is determined by the phase-space
Lagrangian characteristics. The motion of these particles, as hey
are pushed along erratic trajectories by fluctuations of the
fluid-field gradients (in particular, characterizing the fluid
pressure and temperature), is fundamental to transport and mixing
processes in fluids. It is well known that the interaction between
(deterministic and/or turbulent) fluctuations of the fluid fields
and these particles still escapes a consistent theoretical
description. Being a subject of major importance for many
environmental, geophysical and industrial applications, the issue
deserves a careful investigation. A key aspect of fluid dynamics
is the correct definition of the (phase-space) Lagrangian dynamics
which characterizes incompressible fluids. The customary approach
to the Lagrangian formulation is based typically on a
configuration-space description, i.e., on the
introduction of the configuration-space Lagrangian characteristics $\mathbf{r%
}(t),$ spanning the fluid domain $\Omega .$ Here $\mathbf{r}(t)$
denotes the
solution of the initial-value problem:%
\begin{equation}
\left\{
\begin{array}{c}
\frac{D\mathbf{r}}{Dt}=\mathbf{V}(\mathbf{r},t), \\
\mathbf{r}(t_{o})=\mathbf{r}_{o},%
\end{array}%
\right.
\end{equation}%
with $\mathbf{r}_{o}$ an arbitrary vector belonging to the closure $%
\overline{\Omega }$ of $\Omega $ and $\mathbf{V}(\mathbf{r},t)$
being the velocity fluid field, to be assumed continuous in
$\overline{\Omega }$ and suitably smooth in $\Omega $. \ In a
previous paper \cite{Tessarotto2008} IKT has been proven to
advance in time self-consistently the fluid fields, i.e., in such
a way that they satisfy identically the requires set of fluid
equations. For isothermal fluids, this conclusion is consistent
with the results indicated elsewhere by Tessarotto \textit{et
al.}\cite{Ellero2005}. In particular, basic feature of the IKT
approach is to permit to advance in time the fluid fields only by
means of suitable evolution equations, without requiring the
(numerical) solution of the Poisson equation for the fluid
pressure. The purpose of this paper is to point out that, based
the same approach \cite{Tessarotto2008}) for incompressible
thermofluids, \emph{\ the dynamics of tracer particles, of finite
mass (}$m_{T}$\emph{), which are injected in the fluid with
arbitrary velocity, can be rigorously established. }

\section{The exact frictionless conditional dynamics of tracer particles}

In this Section we intend to show that the equations of motion for
a tracer particle in an incompressible fluid can be uniquely
specified, once the velocity probability density are prescribed
(\textit{conditional dynamics}). For definiteness in the sequel we
shall consider an incompressible
thermofluid, described by the fluid fields $\left\{ \rho =\rho _{o}>0,%
\mathbf{V},p\geq 0,T>0\right\} ,${\ denoting respectively the
(constant) mass density, fluid velocity, pressure and temperature
describing the fluid. The fluid fields are assumed to be suitably
smooth, classical solutions of
the incompressible Navier-Stokes-Fourier equations\ (INSFE problem, }\cite%
{Tessarotto2008}), all defined in a bounded connected set (fluid domain) $%
\Omega \subseteq
\mathbb{R}
^{3}$ (with $\overline{\Omega }$ denoting its closure set where
the mass density $\rho $ is strictly positive):
\begin{eqnarray}
&&\left. \nabla \cdot \mathbf{V}=0,\right.   \label{1} \\
&&\left. \frac{D}{Dt}\mathbf{V}=-\frac{1}{\rho _{o}}\left[ \nabla p-\mathbf{f%
}\right] +\nu \nabla ^{2}\mathbf{V},\right.   \label{2} \\
&&\left. \frac{D}{Dt}T=\chi \nabla ^{2}T+\frac{\nu }{2c_{p}}\left( \frac{%
\partial V_{i}}{\partial x_{k}}+\frac{\partial V_{k}}{\partial x_{i}}\right)
^{2}+\frac{1}{\rho _{o}c_{p}}J\equiv K.\right.   \label{4}
\end{eqnarray}%
{\ In standard notation, }$\frac{D}{Dt}\mathbf{V}$ denotes the
fluid
acceleration, with $\frac{D}{Dt}=\frac{\partial }{\partial t}+\mathbf{V}%
\boldsymbol{\cdot \nabla }$ the convective derivative, $J$ is the
quantity of heat generated by external sources per unit volume and
unit time and finally $\mathbf{f}$ is {the force density for which
the} Boussinesq
approximation\ is invoked. Hence, $\mathbf{f}$ is taken on the form{\ }$%
\mathbf{f}=\rho _{o}\mathbf{g}\left( 1-k_{\rho }T\right)
+\mathbf{f}_{1},$ where the first term represents the
(temperature-dependent) gravitational force density and the
second\ one ($\mathbf{f}_{1}$) the action of a possible
non-gravitational externally-produced force density. {As a
consequence, in such a case the force density }$\mathbf{f}$ {reads }$\mathbf{%
f}=\rho _{o}\mathbf{g}\left( 1-k_{\rho }T\right) +\mathbf{f}_{1},$
where the first term represents the (temperature-dependent)
gravitational force density, while the second\ one
($\mathbf{f}_{1}$) the action of a possible non-gravitational
externally-produced force. Hence $\mathbf{f}$ can be written also
as $\mathbf{f}=-\nabla \phi +\mathbf{f}_{R},$ where $\phi =\rho
_{o}gz$ and $\mathbf{f}_{R}=-\rho _{o}\mathbf{g}k_{\rho
}T+\mathbf{f}_{1}$ denote\ respectively the gravitational
potential (hydrostatic pressure) and
the non-potential force density.  Finally, $\nu ${$,$}$\chi ,c_{p}$ and $%
k_{\rho }$ are all real positive constants which denote,
respectively, the kinematic viscosity, the thermometric
conductivity, the specific heat at constant pressure, and the
density thermal-dilatation coefficient. A remarkable aspect of
fluid dynamics is related to the construction of IKT's for
hydrodynamic equations in which the fluid fields are identified
with suitable moments of an appropriate kinetic probability
distribution. This is
achieved introducing a phase-space classical dynamical system (CDS)%
\begin{equation}
\mathbf{x}_{o}\rightarrow \mathbf{x}(t)=T_{t,t_{o}}\mathbf{x}_{o},
\label{classical dynamical system}
\end{equation}%
which uniquely advances in time the fluid fields by means of an
appropriate evolution operator $T_{t,t_{o}}$
\cite{Tessarotto2006,Tessarotto2007}. The CDS is assumed to be
generated by the initial-value problem associate to the
\emph{Lagrangian equations}
\begin{equation}
\left\{
\begin{array}{c}
\frac{d}{dt}\mathbf{r}(t)=\mathbf{v}(t), \\
\frac{d}{dt}\mathbf{v}(t)=\mathbf{F}(\mathbf{r}(t),t;f), \\
\mathbf{r}(t_{o})=\mathbf{r}_{o}, \\
\mathbf{v}(t_{o})=\mathbf{v}_{o},%
\end{array}%
\right.   \label{Lagrangian equations}
\end{equation}%
where $\mathbf{x}=(\mathbf{r,v})\in \Gamma $ and respectively
$\mathbf{r}$ and $\mathbf{v}$ denote an appropriate state-vector
and the corresponding
configuration and velocity vectors.\ Moreover $\mathbf{X}(\mathbf{x}%
,t)\equiv \left\{ \mathbf{v,\mathbf{F}(\mathbf{x},}t;f)\right\} $
is a suitably smooth vector fields, with
$\mathbf{F}(\mathbf{x},t;f)$
representing an acceleration field. \emph{It is assumed that generally} $%
\mathbf{F}(\mathbf{x},t;f)$ \emph{can depend functionally on} $f\equiv f(%
\mathbf{x,}t),$ \emph{to be identified with a suitable probability
density function}$\ $\emph{(pdf)}. In particular, let us require
that $\mathbf{x}$
spans the phase-space $\Gamma =\Omega \times V$, where $V=%
\mathbb{R}
^{3}$ denotes the velocity space. Therefore, introducing the
corresponding velocity pdf $f(\mathbf{x,}t),$ defined so that
\begin{equation}
\int\limits_{V}d^{3}vf(\mathbf{x,}t)=1,  \label{normalization}
\end{equation}%
in $\Gamma $ it fulfills necessarily the integral Liouville
equation

\begin{equation}
J(t)f(\mathbf{x}(t),t)=f(\mathbf{x}_{o},t_{o}). \label{integral
Liouviulle eq}
\end{equation}%
Here, $\mathbf{x}(t)$ denotes the Lagrangian path determined by the CDS (\ref%
{classical dynamical system}), which is assumed to be a
diffeomeorphism at least of class $C^{(2)}(\Gamma \times I\times
I),$ with $I\subset
\mathbb{R}
${\ denoting an appropriate finite time interval}, and
$J(t)=\left\vert \frac{\partial \mathbf{x}(t)}{\partial
\mathbf{x}_{o}}\right\vert $ is corresponding Jacobian{. Then, if
the initial pdf }$f(\mathbf{x}_{o},t_{o})$ is suitably smooth, it
follows that the time-evolved pdf $f(\mathbf{x,}t)$ satisfies
necessarily the differential Liouville equation
\begin{equation}
Lf(\mathbf{x},t)=0,  \label{Liouville}
\end{equation}%
where $L$ denotes the Liouville streaming operator. This equation (\emph{%
inverse kinetic equation}), which may be interpreted as a
Vlasov-type kinetic equation, can in principle be defined in such
a way to satisfy appropriate constraint equations. In particular,
thanks to the arbitrariness of the dynamical systems
(\ref{classical dynamical system}), the velocity moments of
$f(\mathbf{x,}t)$ can be identified so that suitable velocity
moments\ of $f$ coincide with the relevant fluid fields which
characterize an incompressible thermofluid. In particular,
introducing the kinetic pressure $p_{1},$ defined so that
\begin{equation}
p_{1}=p_{0}(t)+p-\phi +\rho _{o}T/m,  \label{ac-6}
\end{equation}%
we can identify $m$ with the average molecular mass of the
classical molecules forming the fluid and the ratio $\rho
_{o}/m\equiv n_{o}$ with the fluid number density $n_{o}$ of the
fluid. In such a case, imposing the
constraints $\mathbf{V,}p_{1}\mathbf{=}\int\limits_{V}d^{3}\mathbf{v}G(%
\mathbf{x,}t)f(\mathbf{x,}t)$, respectively for $G(\mathbf{x,}t)=\mathbf{%
v,\rho }_{o}(\mathbf{v-V})^{2}/3,$ and\emph{\ }subject to the
requirement of strict positivity and regularity for
$f(\mathbf{x,}t),$ the correct form of the acceleration field
$\mathbf{F\equiv F}(\mathbf{x},t;f)$ can be proven to
be \cite{Tessarotto2008}:%
\begin{equation}
\mathbf{F}(\mathbf{x},t;f)=\mathbf{F}_{0}+\mathbf{F}_{1},
\label{ac-1}
\end{equation}%
where\emph{\ }$\mathbf{F}_{0}\mathbf{,F}_{1}$\emph{\ }read respectively%
\begin{equation}
\mathbf{F}_{0}\mathbf{(x,}t;f)=\frac{1}{\rho _{o}}\left[
\mathbf{\nabla
\cdot }\underline{\underline{{\Pi }}}-\mathbf{\nabla }p_{1}+\mathbf{f}_{R}%
\right] +\mathbf{u}\cdot \nabla \mathbf{V+}\nu \nabla
^{2}\mathbf{V,} \label{ac-2}
\end{equation}%
\begin{equation}
\mathbf{F}_{1}\mathbf{(x,}t;f)=\frac{1}{2}\mathbf{u}\left\{ \frac{1}{p_{1}}A%
\mathbf{+}\frac{1}{p_{1}}\mathbf{\nabla \cdot
Q}-\frac{1}{p_{1}^{2}}\left[
\mathbf{\nabla \cdot }\underline{\underline{\Pi }}\right] \mathbf{\cdot Q}%
\right\} +\frac{v_{th}^{2}}{2p_{1}}\mathbf{\nabla \cdot }\underline{%
\underline{\Pi }}\left\{
\frac{u^{2}}{v_{th}^{2}}-\frac{3}{2}\right\} , \label{ac-3}
\end{equation}%
and\emph{\ }%
\begin{equation}
A-\frac{\partial }{\partial t}\left( p_{0}+p\right)
-\mathbf{V\cdot }\left[ \frac{D}{Dt}\mathbf{V-}\frac{1}{\rho
_{o}}\mathbf{f}_{R}\mathbf{-}\nu \nabla ^{2}\mathbf{V}\right]
+\frac{\rho _{o}K}{m},  \label{ac-4}
\end{equation}

\begin{equation}
K=\chi \nabla ^{2}T+\frac{\nu }{2c_{p}}\left( \frac{\partial
V_{i}}{\partial x_{k}}+\frac{\partial V_{k}}{\partial
x_{i}}\right) ^{2}+\frac{1}{\rho _{o}c_{p}}J.  \label{ac-5}
\end{equation}%
Here the notation is given in accordance with Ref.
\cite{Tessarotto2008}.
Thus, $p_{0}(t)$ (to be denoted as \emph{pseudo-pressure}$\mathbb{\ }$\cite%
{Ellero2005}) is a suitably-defined, strictly positive and smooth
real function, to be determined to assure the validity of an
H-theorem (and hence
the strict positivity of $f(\mathbf{x,}t),$ $\mathbf{Q}$ and $\underline{%
\underline{{\Pi }}}$ denote the additional velocity moments
$\mathbf{Q}=\rho
_{o}\int d^{3}v\mathbf{u}\frac{u^{2}}{3}f$ and $\underline{\underline{{\Pi }}%
}=\rho _{o}\int d^{3}v\mathbf{uu}f,$

\emph{The following consequences are implied by the previous
Lagrangian
equations (\ref{Lagrangian equations}) [together with the definitions (\ref%
{ac-1})-(\ref{ac-6})]:}

\begin{enumerate}

\item the Lagrangian equations (\ref{Lagrangian equations}) can be interpreted either
as deterministic or stochastic, depending whether the fluid fields
themselves are treated as such (see also
Ref.\cite{Tessarotto2008-aa});

\item $\mathbf{F}(\mathbf{x},t;f)$ can be interpreted as the "conditional"
Lagrangian acceleration, which depends functionally on the form of the pdf $%
f(\mathbf{x,}t)$. In particular, in the case of a turbulent fluid
also the pdf must be considered as stocastic
\cite{Tessarotto2008-aa};

\item it is possible to prove that the functional form of the vector field $%
\mathbf{F}(\mathbf{x},t;f)$ can be uniquely specified once
$f(\mathbf{x,}t)$ \ is prescribed$;$

\item the form of $f(\mathbf{x,}t)$ is determined uniquely by its initial
condition \ $f(\mathbf{x}_{o},t_{o})$ [see Eq.(\ref{integral Liouviulle eq}%
)] and by the time-evolution operator $T_{t,t_{o}};$

\item in turn, $T_{t,t_{o}}$ is defined uniquely by the CDC (\ref{classical
dynamical system}) $f(\mathbf{x,}t);$

\item a particular solution of Eq.(\ref{Liouville}) is delivered by the
Maxwellian pdf
\begin{equation}
f_{M}(\mathbf{x},t)=\frac{1}{\pi ^{2}v_{th,p}^{3}}\exp \left\{ -\frac{u^{2}}{%
v_{th,p}^{2}}\right\} ,  \label{Maxwellian}
\end{equation}%
where $\mathbf{u}=\mathbf{v}-\mathbf{V}(\mathbf{r},t)$ and $v_{th,p}=\sqrt{%
2p_{1}(\mathbf{r},t)/\rho }$ are respectively the relative and the
thermal velocities.

\item the case of an isothermal fluid is recovered by imposing that there
results identically $T(\mathbf{r},t)=const.$ and $K\equiv 0$ [see
the r.h.s. of Eq.(\ref{4})].
\end{enumerate}

We conclude that\emph{, the Lagrangian equations (\ref{Lagrangian equations}%
), when letting }$m=m_{T},$ \emph{\ define the equations of motion
for a
tracer particle with initial conditions} $\left\{ \mathbf{r}(t_{o}),\mathbf{v%
}(t_{o})\right\} =\left( \mathbf{r}_{o},\mathbf{v}_{o}\right) $
\emph{and subject to the conditional acceleration}
$\mathbf{F}(\mathbf{x},t;f).$

\section{Conclusion and outlook}

In this Note we have proven that the exact dynamics of tracer
particles injected in incompressible thermofluids can be
rigorously established. The result appears relevant not only from
the theoretical viewpoint but also for its applications and in
particular for the description and monitoring of pollutant
dynamics in the atmosphere.

\ We stress that only weak restrictions have been assumed on the
form of the
pdf, affecting the form of the conditional acceleration $\mathbf{F}(\mathbf{x%
},t;f).$ In addition, the theory applies in principle to arbitrary
fluid fields which are strong solutions of the INSFE problem.
Therefore, this permits in principle also the proper treatment of
turbulent, i.e., stochastic fields. As a consequence, the
formulation here presented is susceptible of applications in
turbulence theory.



\section*{Acknowledgments}
Work developed in cooperation with the CMFD Team, Consortium for
Magneto-fluid-dynamics (Trieste University, Trieste, Italy). \
Research developed in the framework of the MIUR (Italian Ministry
of University and Research) PRIN Programme: \textit{Modelli della
teoria cinetica matematica nello studio dei sistemi complessi
nelle scienze applicate}. The support COST Action P17 (EPM,
\textit{Electromagnetic Processing of Materials}) and GNFM
(National Group of Mathematical Physics) of INDAM (Italian
National Institute for Advanced Mathematics) is acknowledged.

\section*{Notice}
$^{\S }$ contributed paper at RGD26 (Kyoto, Japan, July 2008).
\newpage



\end{document}